\documentclass[11pt]{article}
\usepackage{amssymb}
\usepackage{amsmath}
\usepackage{color}
\usepackage{verbatim}
\usepackage[colorlinks,linkcolor=blue,citecolor=green]{hyperref}
\newtheorem{theorem}{Theorem}

\newenvironment{proof}[1][Proof]{\textbf{#1.} }{\ \rule{0.5em}{0.5em}}

\usepackage{graphicx}
\allowdisplaybreaks[3]

\title{A reduction procedure for determining \\ exact solutions of second order hyperbolic equations}
\author{Natale Manganaro$^1$, \quad Alessandra Rizzo$^1$\\
\\
 \small  $^1$ Department of Mathematical, Computer,\\
  \small Physical and Earth Sciences (MIFT)\\
  \small University of Messina, \\
  \small V.le F. Stagno D'Alcontres 31, I-98166 Messina, Italy \\
\small e.mail: nmanganaro@unime.it, alerizzo@unime.it }
\date{}

\begin{document}
\maketitle
\begin{abstract}
In this paper we develop a systematic reduction procedure for determining intermediate integrals of second order hyperbolic equations so that exact solutions of the second order PDEs under interest can be obtained by solving first order PDEs. We give some conditions in order that such a procedure holds and, in particular,  we characterize classes of linear second order hyperbolic equations for which the general solution can be found.   
\end{abstract}

\vspace{0.3cm}

\vspace{0.3cm}

\section{Introduction}
Determining exact solutions of partial differential equations (PDEs) is of great interest not only for its theoretical value but also for possible applications. To this end, along the years, many mathematical approaches have been proposed (see \cite{mel1} for an exhaustive review of such methods). In particular, in $1964$ N. N. Yanenko proposed the method of differential constraints  and applied it to gas-dynamics \cite{jan}. The main idea is to add to the governing equations one or more further differential relations and to look for solutions of the full system. In such a case the added equations  play the role of constraints because they select the class of exact solutions of the original system we are looking for. The first step of such a method is to study the compatibility between the original equations and the constraints. Usually to develop such an analysis is a hard task. For this reason further assumptions were made in the reduction algorithm and many contributions were given in the litterature \cite{fsy}-\cite{rv}.

Within such a theoretical framework, the method of the intermediate integrals permits, in principle, to obtain particular exact solutions of higher order PDEs. In fact, we consider the second order equation
\begin{equation}
F\left(x_i, u, \frac{\partial u}{\partial x_j}, \frac{\partial^2 u}{\partial x_i x_j}\right)=0, \quad \quad i,j=1,...,n \label{H}
\end{equation} 
and we add to it the first order equation
\begin{equation}
f\left(x_i, u, \frac{\partial u}{\partial x_j} \right)=0, \quad \quad i,j=1,...,n. \label{hv}
\end{equation}  
The equation (\ref{hv}) is called intermediate integral of (\ref{H}) if all solutions of (\ref{hv}) are also solutions of (\ref{H}). Therefore, in such a case, in order to find particular exact solutions of the original second order equation (\ref{H}) we are led to solve the first order equation (\ref{hv}) which, in fact, plays the role of constraint. In this sense the method of differential constraints could be considered as a generalization of the method of the intermediate integrals to systems of PDEs. The interested reader can find more details about such an approach in \cite{mel1} where the procedure is also developed for a linear second order hyperbolic equation. 

Proving the integrability of PDEs systems is of great interest not only from a theoretical point of view \cite{fv}. Within such a framework, quite recently, in \cite{mrv} a class of one-dimensional  hyperbolic wave equations with non constant speed admitting intemediate integrals have been characterized. In particular, for some special forms of the wave speed, the general solution of the wave equation has been obtained and, in turn, the integrability of a class of Hamiltonian systems has been proved. 

Following the analysis developed in \cite{mrv}, the main aim of this paper is to look for intermediate integrals of second order hyperbolic equations and, consequently, to determine exact solutions of such equations. In particular, in section $2$ we develop a procedure for characterizing intemediate integrals for a generic second order hyperbolic equation. In section $3$ some examples of hyperbolic equations for which the approach developed previously has been useful for characterizing exact solutions are given. In section $4$ we apply our procedure in order to determine the general solution of linear second order hyperbolic equations. Finally, in section $5$ some final remarks and comments are given.
 
\section{Reduction procedure}
The main aim of this section is to provide intermediate integrals for second order hyperbolic equations. Therefore, let us consider the hyperbolic equation
\begin{equation}
u_{tt}-a^2 (x,t,u) u_{xx}=f(x,t,u,u_x, u_t) \label{he}
\end{equation}
where $a(x,t,u)$ is the wave speed and $f(x,t,u,u_x, u_t)$ is a given function. We append to (\ref{he}) the constraint
\begin{equation}
F(x,t,u,u_x, u_t)=0. \label{con}
\end{equation}
We are able to prove the following
 \begin{theorem}
The relation (\ref{con}) is an intermediate integral of the equation (\ref{he}) if $F$ assumes the form
\begin{equation}
F=u_t-\lambda u_x-g(x,t,u) \label{fc}
\end{equation}
 and the following condition is satisfied
\begin{equation}
\left(\lambda_t+\lambda \lambda_x+2\lambda g_u+g\lambda_u \right)u_x+2\lambda \lambda_u u_{x}^{2}+g_t+\lambda g_x+gg_u=f \label{con1}
\end{equation}
where $\lambda=\mp a(x,t,u)$ and $g(x,t,u)$ is a function to be determined.
\end{theorem}

\noindent
\begin{proof}
 Setting $u_x=q$, $u_t=p$, $u_{xx}=z$, $u_{tt}=r$ and $u_{xt}=s$,  in order to study the compatibility between (\ref{con}) and (\ref{he}), we differentiate (\ref{con}) so that, taking (\ref{he}) into account, we have the following linear system in $z$, $r$ and $s$
\begin{eqnarray}
&&F_p s+ F_q z=-F_x -q F_u \label{r1} \\
&&F_p r+F_q s=-F_t -p F_u \label{r2} \\
&&r-a^2 z=f \label{r3}
\end{eqnarray}
The solution of the equation (\ref{fc}) is obtained in terms of one arbitary function. Therefore (\ref{fc}) is an intermediate integral of (\ref{he}) only if the solution of (\ref{he}) has the same arbitrariness so that we are led to require that 
\begin{equation}
det \begin{Vmatrix}
F_p & F_q & 0 \\
F_q & 0 & F_p \\
0 & -a^2 & 1
\end{Vmatrix}=0.   \label{det}
\end{equation}
In fact, if the system (\ref{r1})-(\ref{r3}) admits one and only one solution (i. e. if the second order derivatives of $u$ can be calculated univocally from (\ref{r1})-(\ref{r3})), then the solution of (\ref{he}) should be given in terms of arbitrary constants. Condition (\ref{det}) gives
$$
F_p-\lambda F_q=0, \quad \quad \lambda=\pm a
$$ 
whose integration leads to
\begin{equation}
F=p-\lambda q - g(x,t,u) \label{F}
\end{equation}
Finally, by substitution (\ref{F}) into (\ref{r1})-(\ref{r3}), we obtain the condition (\ref{con1}).
\end{proof}

\vspace{0.3cm}
\noindent
{\bf Remark 1.} Owing to Theorem 1, since the relation (\ref{con}) is an intermediate integral only if $F$ assumes the form (\ref{fc}), in order to determine exact solutions of (\ref{he}), we have to integrate the first order equation
\begin{equation}
u_t-\lambda(x,t,u) u_x=g(x,t,u)  \label{pde}
\end{equation}
where the function $g$ must be determined according to (\ref{con1}). Therefore, the solution of (\ref{he}) will be given in terms of one arbitrary function. We will show later that, in some cases, the procedure here developed leads to obtain the solution in terms of two arbitrary functions (i. e. we find the general solution of the equation under interest). This happens, for instance, when the second order equation is linear.

\vspace{0.3cm}
\noindent
{\bf Remark 2.} As consequence of Theorem 1, owing to condition (\ref{con1}) and taking (\ref{pde}) into account, we notice that the function $f$ involved in the equation (\ref{he}) can adopt one of the following forms
\begin{eqnarray}
&&f=A_0 + B_0 u_x + C_0 u_{x}^{2}   \label{f0} \\
&&f=A_1+B_1 u_t+C_1 u_{x}^{2}   \label{f1} \\
&&f=A_2+B_2 u_t+C_t u_{t}^{2} \label{f2} \\
&&f=A_3+B_3 u_t + C_3 u_x u_t +D_3 u_x \label{f3} \\
&&f=A_4 + B_4 u_x + C_4 u_t + D_4 u_{t}^{2} \label{f4} \\
&&f=A_5 +B_5 u_x + C_5 u_x u_t \label{f5}
\end{eqnarray} 
where $A_i (x,t,u)$, $B_i (x,t,u)$, $C_i (x,t,u)$ and $D_i (x,t,u)$ are suitable functions given in terms of the coefficients involved in the equation at hand. Therefore, the class of second order hyperbolic equations admitting intermediate integrals is given by (\ref{he}) supplemented by $f$ defined according to one of the forms (\ref{f0})-(\ref{f5}).

\vspace{0.3cm}
\noindent
{\bf Remark 3.} Owing to Theorem $1$, since $\lambda=\mp a(x,t,u)$, from (\ref{pde}) we have two intermediate integrals of (\ref{he}). Thus, two different particular solutions of (\ref{he}) are obtained by integrating the two reductions (\ref{pde}).

\vspace{0.3cm}
\noindent 
In the next section, in order to illustrate better the approach here developed, we will give some examples  for which such a procedure has been useful for determining  exact solutions of equations belonging to the class (\ref{he}).

\section{Intermediate integrals and exact solutions}
The key point of the reduction procedure described in the previous section is the condition (\ref{con1}). In fact, once the function $f$ is specified, we have to require that relation (\ref{con1}) is satisfied for all  solutions of (\ref{pde}) (i. e. $\forall \, u_x$). Such a requirement leads to a set of compatibility conditions involving the coefficients of the equation at hand as well as the unknown function $g(x,t,u)$. Once the compatibility conditions are solved and, in turn, $g$ is determined, we can integrate (\ref{pde}).

\vspace{0.2cm}
\noindent
$1$. As first example we consider the equation
\begin{equation}
u_{tt}-a^2 (u) u_{xx}=2 a a^\prime u_{x}^{2}+\Phi (u) u_x +h(x,t)+q(u) \label{e1}
\end{equation}
which was widely studied in the literature. In fact when $\Phi=0$ and $h=0$ or when $q=0$ and $h=0$ equation (\ref{e1}) has been considered in \cite{huang} while when $\Phi=0$ and $q=0$ it was studied in \cite{zhou}.   In the present case the function $ f$ adopts the form
$$
f=2a a^\prime u_{x}^{2}+\Phi (u)u_x +h(x,t)+q(u)
$$
so that from (\ref{con1}) the following compatibility conditions are obtained
\begin{eqnarray}
&&2\lambda g_u+ \lambda_u g=\Phi (u) \label{g11}  \\
&&g_t + \lambda g_x +g g_u =h(x,t) + q(u) \label{g12}
\end{eqnarray}
where $\lambda =\pm a(u)$. In what follows we consider the case $\lambda=a$ because a similar reduction can be obtained when $\lambda=-a$ (see Remark 3).

Thus, integration of (\ref{g11}) leads to
\begin{equation}
g=\frac{1}{\sqrt{a}}\left( \varphi(u)+ G(x,t) \right), \quad \varphi=\int{\frac{\Phi (u)}{2\sqrt{a}}\, du}  \label{G1}
\end{equation}
where $G(x,t)$ is a function to be specified. By substitution of (\ref{G1}) in (\ref{g12}), after some calculations, we get the following two cases

\vspace{0.2cm}
$1.1)$ If $G \neq const.$, we have
\begin{eqnarray}
&&\frac{d}{du}\left( \frac{1}{\sqrt{a}}\right)=\beta_0 \sqrt{a}-\alpha_0 a-\gamma_0 \label{pp1} \\
&&\frac{d}{du}\left( \frac{\varphi}{\sqrt{a}}\right)+\varphi \frac{d}{du} \left( \frac{1}{\sqrt{a}}\right)=\beta_1 \sqrt{a}-\alpha_1 a-\gamma_1 \label{p2} \\
&& q\sqrt{a}=\varphi \frac{d}{du}\left( \frac{\varphi}{\sqrt{a}}\right)-\beta_2 \sqrt{a}+\alpha_2 a+\gamma_2 \label{p3} \\
&&h=\beta_0 G^2 +\beta_1 G+\beta_2 \label{p4}
\end{eqnarray}
where $\alpha_i$, $\beta_i$, $\gamma_i$ are constants. Furthermore $G(x,t)$ must satisfy the relations
\begin{eqnarray}
&&G_x=\alpha_0 G^2 +\alpha_1 G+\alpha_2 \label{qq1} \\
&&G_t=\gamma_0 G^2 +\gamma_1 G+\gamma_2 \label{qq2}
\end{eqnarray}
whose compatibility conditions require
\begin{equation} 
\alpha_0 \gamma_1=\alpha_1 \gamma_0, \quad \alpha_0 \gamma_2=\alpha_2 \gamma_0, \quad \alpha_1 \gamma_2=\alpha_2 \gamma_1. \label{comp}
\end{equation}

\vspace{0.2cm}
$1.2)$ If $G=k_0=const.$, we find
\begin{equation}
h=h_0=const. , \quad \quad q=\frac{1}{2}\frac{d}{du}\left( \frac{\varphi +k_0}{\sqrt{a}}\right)^2 \label{cc1}
\end{equation}
while $a(u)$ is unspecified. 

Therefore, once $a(u)$, $q(u)$, $\Phi(u)$, $h(x,t)$ are assigned according to (\ref{pp1})-(\ref{p4}) or to (\ref{cc1}), exact solutions of (\ref{e1}) can be obtained by solving equation (\ref{pde}) supplemented by (\ref{G1}) where $G$ must be calculated from (\ref{qq1}), (\ref{qq2}) (in the case $1.1)$) or $G=k_0$ (in the case $1.2)$).  In what follows we consider the three model equations arising from (\ref{e1}) when $q=\Phi=0$ or $q=h=0$ or $h=\Phi=0$.

\vspace{0.2cm}
\noindent
{\it i)} When $q=\Phi=0$, it is simple matter to verify that from (\ref{p2})-(\ref{p4}) and (\ref{qq1}), (\ref{qq2}) we obtain
\begin{equation}
g=-\frac{1}{\sqrt{a}\left( \alpha_0 x + \gamma_0 t \right)}, \quad \quad h=\frac{\beta_0}{\left( \alpha_0 x+ \gamma_0 t \right)^2} \label{ca1}
\end{equation}
while $a(u)$ must be given according to (\ref{pp1}). Therefore, taking (\ref{ca1}) into account, integration of (\ref{pde}) will give an exact solution of (\ref{ca1}) parameterized by one arbitrary function. For instance, if we assume $\alpha_0=0$, from (\ref{pde}) we find
$$
\int_{u_0(\sigma)}^{u}{\sqrt{a(u)}\, du}=-\frac{1}{\gamma_0} \ln{\left( \frac{t}{t_0}\right)}, \quad \quad x=-\int_{t_0}^{t}{a\left( u \left( \sigma , t \right)\right) \, dt} + \sigma
$$
where $t_0$ is a constant while $u_0 (\sigma)$ denotes an arbitrary function. 

As far as the case $1.2)$ is concerned, it is simple matter to verify that it is not compatible with the assumption $q=\Phi=0$ unless $a=const.$

\vspace{0.2cm}
\noindent
{\it ii)} When $q=h=0$, we find that the case $1.2)$ leads to 
\begin{equation}
g=k_1, \quad \quad \Phi=k_1  a^\prime (u)  \label{ca2}
\end{equation}
where $k_1$ is an arbitrary constant. Therefore, integration of (\ref{pde}) gives
\begin{equation}
u=k_1 t+ u_0 (\sigma), \quad \quad x=-\int_{0}^{t}{a\left(u \left(t, \sigma \right) \right) \, dt} + \sigma \label{sss1}
\end{equation} 
where $u_0(\sigma)$ is an arbitrary function. Relation (\ref{sss1}) characterizes a solution  of (\ref{e1}) (with $q=h=0$, $\Phi=k_1 a^\prime (u)$ and $a(u)$ unspecified) in terms of one arbitrary function. Moreover we notice that, owing to (\ref{ca2})$_2$, the equation (\ref{e1}) is equivalent  to the first order system
\begin{eqnarray}
&&u_t - v_x=0 \nonumber \\
&&v_t-a^2 (u) u_x=k_1 a(u) \nonumber
\end{eqnarray}
which is the well known non-homogeneous p-system. Furthermore it is not difficult to ascertain that the case $1.1)$ is consistent with the procedure here developed only if $a=const.$

\vspace{0.2cm}
\noindent
{\it iii)} When $h=\Phi=0$, in the case $1.2)$ we deduce  that
\begin{equation}
g=\frac{k_0}{\sqrt{a}}, \quad \quad q=\frac{k_0^2}{2} \frac{d}{du}\left( \frac{1}{a(u)}\right) \label{v15}
\end{equation}
while $a(u)$ is unspecified. In such a case, owing to (\ref{v15})$_1$, from (\ref{pde}) we have
\begin{equation}
\int_{u_0(\sigma)}^{u}{\sqrt{a(u)} \, du}=k_0 t, \quad \quad x=-\int_{0}^{t}{a\left( u\left( \sigma , t \right)\right) \, dt} + \sigma \label{er1}
\end{equation}
where $u_0(\sigma)$ is an arbitrary function. Therefore, once $a(u)$ is given, relation (\ref{er1}) gives a solution of (\ref{e1}) supplemented by (\ref{v15})$_2$.

Furthermore, in the case $1.1)$ we obtain
\begin{equation}
q=\frac{\gamma_2}{\sqrt{a}}+\alpha_2 \sqrt{a}, \quad \quad g=\frac{G(x,t)}{\sqrt{a}} \label{v1}
\end{equation}
where, assuming $a(u) \neq const.$ (i. e. $\alpha_0^2+\gamma_0^2 \neq 0$), owing to (\ref{qq1}), (\ref{qq2}) we find
\begin{eqnarray}
&&G=c_0 \tan{\left(c_0 \sigma +c_1\right)} \hspace{1.5cm} \mbox{if $\frac{\alpha_2}{\alpha_0}=c_0^2$}   \label{GG1} \\
&&G=c_0 \frac{1+e^{2c_0 \sigma + c_1}}{1-e^{2c_0 \sigma +c_1}} \hspace{1.8cm} \mbox{if $\frac{\alpha_2}{\alpha_0}=-c_0^2$} \label{GG2} \\
&&G=-\frac{1}{\sigma +c_1} \hspace{2.8cm} \mbox{if $\alpha_2 =\gamma_2=0$} \label{GG3} 
\end{eqnarray}
while, from (\ref{pp1}) we obtain
\begin{eqnarray}
&&\frac{1}{\sqrt{a}}-c_2 \arctan{\left( \frac{1}{c_2 \sqrt{a}}\right)}=-\gamma_0 u \quad \quad \mbox{if $\gamma_0 \neq 0$ and $\frac{\alpha_0}{\gamma_0}=c_2^2$} \label{A1} \\
&&\frac{1}{\sqrt{a}}-\frac{c_2}{2} \ln{\left( \frac{1+c_2 \sqrt{a}}{1-c_2 \sqrt{a}} \right)}=-\gamma_0 u \quad \quad \mbox{if $\gamma_0 \neq 0$ and $\frac{\alpha_0}{\gamma_0}=-c_2^2$} \label{A2} \\
&&a=\frac{1}{\gamma_0^2 u^2} \quad \; \quad \quad \quad \quad \quad \quad \quad \quad \quad \quad \quad \mbox{if $\alpha_0=0$} \label{A3} \\
&&a=\left( -3\alpha_0 u\right)^{-\frac{2}{3}} \; \quad \, \quad \quad \quad \quad \quad \quad \quad \quad \mbox{if $\gamma_0=0$} \label{A4}  
\end{eqnarray}
In (\ref{GG1})-(\ref{A4}) $c_i$ are constants while $\sigma=\alpha_0 x+\gamma_0 t$. In such a case, integration of (\ref{pde}) can be accomplished  once $a(u)$ is given according to (\ref{A1})-(\ref{A4}) and taking (\ref{v1})$_2$ into account supplemented by (\ref{GG1})-(\ref{GG3}). For instance if $\alpha_0=\alpha_2=\gamma_2=0$ the functions $a$ and $G$ are given, respectively, by (\ref{A3}) and (\ref{GG3}), so that from (\ref{pde}) the following solution of (\ref{e1}) is obtained
$$
u=\frac{ u_0 (z)}{ t +t_0}, \quad \quad x=-\frac{1}{3  \gamma_0^2 u_0^2 (z)} \left( \left(t+t_0 \right)^3 -t_0^3 \right)+z
$$
where $u_0(z)$ is an arbitrary function and we set $t_0=\frac{c_1}{\gamma_0}$.

As final case, we assume $h=\Phi=q=0$ so that  equation (\ref{e1}) specializes to
\begin{equation}
u_{tt}=\partial_x \left( a^2 (u) u_x \right).  \label{ps}
\end{equation}
which by setting $v_x=u_t$ and $v_t=a^2 u_x$ is equivalent to the homogeneous p-system.  Furthermore, if we require $g=0$,  the compatibility conditions (\ref{g11}), (\ref{g12}) are identically satisfied and from (\ref{pde}) we obtain
\begin{equation}
u=u_0 (\xi), \quad \quad \xi=x +u_0(\xi)  t \label{s31}
\end{equation} 
with $u_0 (\xi)$ denoting an arbitrary function. Therefore, relation (\ref{s31}) gives a solution of (\ref{ps}) $\forall a(u)$.

\vspace{0,2cm}
\noindent
{\bf Remark 4.} We notice that when $a(u)=\frac{1}{u}$, $h=2$, $\Phi = q =0$, then equation (\ref{e1}) specializes to the constant astigmatism equation considered in \cite{mp} where different new solutions of such equation have been obtained. It is simple to verify that the results determined in \cite{mp} can be recovered by means of the more general approach here developed starting from the compatibility conditions (\ref{g11}), (\ref{g12}). 

Furthermore it is also of interest  to remark that a parametric solution of equation (\ref{ps}) depending of two arbitrary functions  was obtained in \cite{kap0} when the coefficient $a(u)$ adopts the form
$$
a=u^{\frac{4n}{1-2n}}
$$
for any integers $n$.

\vspace{0.3cm}
\noindent
$2$. A second example is given by the equation
\begin{equation}
u_{tt}=u^2 u_{xx}-u_t +\frac{2}{u}u_t^2  \label{e5}
\end{equation}
which was considered in \cite{anco}. In the present case we have $\lambda = \mp u$ while 
$$
f=-g+\frac{2}{u}g^2+\left(4g-u\right)u_x+2uu_x^2.
$$
In what follows we will consider the case $\lambda=u$. Therefore, from (\ref{con1}) we obtain
$$
2ug_u-3g=-u, \quad \quad g_t+ug_x+gg_u=-g+\frac{2}{u}g^2
$$
whose integration leads to $g=u$, so that from (\ref{pde}) the following solution of (\ref{e5}) is given
\begin{equation}
u=u_0(\sigma) \, e^t, \quad \quad x=-u_0(\sigma) \left( e^t - 1 \right) + \sigma \label{n}
\end{equation}
where $u_0(\sigma)$ denotes an arbitrary function.

\vspace{0.3cm}
\noindent
$3$. Here, we consider the equation
\begin{equation}
u_{tt}=c^2 u_{xx}+q_1(x)u_t+q_2(x) u \label{e6}
\end{equation}
which was studied in \cite{liu}. It results that $\lambda= \mp c$, with $c=const.$, while
$$
f=q_2 u+q_1 g-\lambda q_1 u_x
$$
so that from (\ref{con1}) we get
\begin{equation}
2g_u=q_1, \quad \quad g_t+\lambda g_x +gg_u=q_1 g+q_2 u. \label{rs}
\end{equation}
Integration of (\ref{rs}) leads to
\begin{eqnarray}
&&g=\frac{q_1 (x)}{2}u + \gamma (x,t) \label{gh1} \\
&&\gamma_t + \lambda \gamma_x=\frac{q_1 (x)}{2} \gamma \label{gh2} \\
&&q_2 (x)=\frac{\lambda}{2}q_1^\prime (x)-\frac{q_1^2 (x)}{4} \label{gh3}
\end{eqnarray}
Therefore, once $q_1(x)$ and $q_2(x)$ are given according to (\ref{gh3}), taking (\ref{gh2}) into account, integration of (\ref{pde}) supplemented by (\ref{gh1}) leads to an exact solution of (\ref{e6}). For instance, if we assume $\gamma=0$, by integration of (\ref{pde}) we have, in the case $\lambda=-c$,
\begin{equation}
u_1=\hat{u}_0(\sigma) e^{\int{\frac{q_1 (x)}{2c}\, dx}}, \quad \quad \sigma=x-ct \label{ll1}
\end{equation}
while, when $\lambda=c$, we obtain
\begin{equation}
u_2=\tilde{u}_0(\xi) e^{-\int{\frac{q_1 (x)}{2c}\, dx}}, \quad \quad \xi=x+ct. \label{ll2}
\end{equation}
where $\hat{u}_0 (\sigma)$ and $\tilde{u}_0 (\xi)$ are arbitrary functions. 

In passing we notice that when $q_1=2k_0$ (where $k_0$ is a constant) so that $q_2=-k_0^2$, equation (\ref{e6}) specializes to the linear telegraph equation studied in \cite{sri}, while when $q_1=-1$ and $q_2=-\frac{1}{4}$, equation (\ref{e6}) is the hyperbolic Cahn-Allen equation with free energy under the form $\epsilon=\left(\frac{u^2}{8M_u}-c\right)$ where $M_u$ denotes the mobility parameter of the order parameter $u$ \cite{niz}. In the next section we will study such linear cases.

\vspace{0.3cm}
\noindent
$4$. Now we put our attention to the equation
\begin{equation}
u_{tt}-u_{xx}=-c_0 u_t +h(x,t,u) \label{e2}
\end{equation}
which has been considered in \cite{alonso}. Since here $f=-c_0 u_t +h$, from (\ref{con1}) we find
\begin{eqnarray}
&&2 g_u=-c_0  \label{s3} \\
&&g_t+\lambda g_x +gg_u=-c_0 g+h \label{s4}
\end{eqnarray}
where $\lambda=\mp 1$. Integration of (\ref{s3}) and (\ref{s4}) gives
\begin{eqnarray}
&&g=-\frac{c_0}{2}u+\gamma(x,t) \label{gg1} \\
&&h=-\frac{c_0^2}{4}u + h_0 (x,t) \label{gg2}
\end{eqnarray}
where $h_0 (x,t)$ is an unspecified function, while $\gamma(x,t)$ must satisfy the equation
\begin{equation}
\gamma_t + \lambda \gamma_x=-\frac{c_0}{2} \gamma +h_0 (x,t). \label{gaa}
\end{equation}
Because of the form of the function $h(x,t)$ given in (\ref{gg2}), the equation (\ref{e2}) assumes a linear form. In the next section we will give the general solution of such linear equation.

\section{General solution for linear equations}
The main aim of this section is to characterize classes of linear second order hyperbolic equations for which the procedure here considered permits to determine their general solution. The idea is based on the fact that, owing to Theorem $1$, the equation (\ref{he}) admits the intermediate integrals
\begin{equation}
u_t \mp a u_x=g^{\pm}(x,t,u) \label{red}
\end{equation}
provided that the condition (\ref{con1}) is satisfied for both reductions. Thus, when the equation (\ref{he}) is linear, its general solution will be given by the linear combination of the solutions of (\ref{red}).  

Here we consider the equation (\ref{he}) with $a(x,t)$ and $f=A(x,t)u_x+H(x,t)u+B(x,t)+G(x,t)u_t$. Taking (\ref{red}) into account, from the compatibility condition (\ref{con1}) we get
\begin{eqnarray}
&&\pm a_t+aa_x \pm 2a g^{\pm}_{u}=A\pm aG \label{aa} \\
&&g^{\pm}_{t}\pm a g^{\pm}_{x}+g^{\pm} g^{\pm}_{u}=Hu+B+Gg^{\pm} \label{gg}
\end{eqnarray}
After some algebra, from (\ref{aa}), (\ref{gg}), we obtain
\begin{equation}
g^{+}=\gamma(x,t)u+\alpha(x,t), \quad \quad g^{-}=\eta(x,t)u+\beta(x,t) \label{a3}
\end{equation}
where
\begin{equation}
\gamma=\frac{1}{2a}\left( A+aG-\left( a_t +aa_x \right) \right), \quad \quad 
\eta=-\frac{1}{2a} \left( A-aG+\left(a_t - a a_x \right) \right) \label{ge}
\end{equation}
while the functions $\alpha(x,t)$ and $\beta(x,t)$ are determined by
\begin{eqnarray}
&& \alpha_t + a \alpha_x =B+ \left( G-\gamma \right) \alpha \label{al} \\
&&\beta_t - a \beta_x=B+\left( G- \eta \right) \beta \label{be}
\end{eqnarray}
Furthermore, the following structural conditions  must be satisfied
\begin{eqnarray}
&&\gamma_t + a \gamma_x =H+G\gamma - \gamma^2 \label{ga} \\
&&\eta_t - a \eta_x=H+G\eta - \eta^2 \label{et}
\end{eqnarray}
Therefore, we are able to give the following
\begin{theorem}
The general solution of the equation
$$
u_{tt}- a^2 (x,t) u_{xx}=A(x,t)u_x + H(x,t)u+B(x,t)+G(x,t) u_t
$$
is given by the linear combination of the solutions of the first order equations (\ref{red}) supplemented by (\ref{a3}), provided that the conditions (\ref{ga}) and (\ref{et}) are satisfied.
\end{theorem}

\vspace{0.2cm}
\noindent
In the following we will give some examples for which such an approach has been useful for determining the general solution of some linear equations.

\vspace{0.2cm}
\noindent
{\it i)}  As first example, we consider the equation (\ref{e6}). From (\ref{ge}) we have
\begin{equation}
\gamma=\eta=\frac{q_1}{2}   \label{vv}
\end{equation}
while from (\ref{ga}) and (\ref{et}) we deduce
\begin{equation}
q_1=const. \quad \quad \mbox{and} \quad \quad q_2=-\frac{q_1^2}{4}.  \label{kk}
\end{equation}
Furthermore, integration of (\ref{al}) and (\ref{be}) gives
\begin{equation}
\alpha=\alpha_0 (\sigma) e^{\frac{q_1}{2}t}, \quad \quad \beta=\beta_0 (\xi) e^{\frac{q_1}{2}t}  \label{p1}
\end{equation}
where
$$
\sigma=x-ct, \quad \quad \xi=x+ct
$$
while $\alpha_0$ and $\beta_0$ are arbitrary functions. Finally, by solving equations (\ref{red}) supplemented by (\ref{a3}) along with (\ref{vv}) and (\ref{p1}), we have
$$
u_1=e^{-\frac{q_1}{4c}\sigma}\left( -e^{\frac{q_1}{4c}\xi} \int{\frac{\alpha_0(\sigma)}{2c} d\sigma} +u_{1}^{0}(\xi)\right), \quad \quad u_2=e^{\frac{q_1}{4c}\xi}\left( e^{-\frac{q_1}{4c}\sigma} \int{\frac{\beta_0(\xi)}{2c} d\xi} +u_{2}^{0}(\sigma)\right) 
$$
where $u_{1}^{0}$, $u_{2}^{0}$ are arbitrary functions. Therefore, the general solution of (\ref{e6}) with (\ref{kk}) is
$$
u=u_1+u_2=u_{1}^{0}(\xi)\, e^{-\frac{q_1}{4c}\sigma}+u_{2}^{0}(\sigma)\,  e^{\frac{q_1}{4c}\xi}
$$
where, without loss of generality, we set $\alpha_0=\beta_0=0$.

\vspace{0.3cm}
\noindent
 {\ ii)} Now we consider the equation
\begin{equation}
u_{tt}-a(x) u_{xx}=a^\prime (x) u_x - c(x) u +h(x,t) \label{e3}
\end{equation}
which was studied in \cite{pani}. In the present case
\begin{equation}
\gamma=\frac{a^{\prime}(x)}{4 \sqrt{a}}, \quad \quad \eta=-\frac{a^{\prime}(x)}{4 \sqrt{a}}   \label{gaet}
\end{equation}
while, from (\ref{ga}) and (\ref{et}) we obtain the condition
\begin{equation}
c(x)=-\frac{a^{\prime \prime}}{4}+\left( \frac{a^\prime}{4 \sqrt{a}}\right)^2. \label{q2}
\end{equation}
Integration of (\ref{al}) and (\ref{be}) leads to
\begin{equation}
\alpha=a^{-\frac{1}{4}} \left( \int{h \left( x, t\left(x, \xi \right)\right)} \, a^{-\frac{1}{4}} \, dx +\alpha_0 (\xi) \right), \quad \beta=a^{-\frac{1}{4}} \left( -\int{h \left( x, t\left(x, \sigma \right)\right)} \, a^{-\frac{1}{4}} \, dx +\beta_0 (\sigma) \right) \label{albe2} 
\end{equation}
where
$$
\xi=t-\int{\frac{dx}{\sqrt{a(x)}}}, \quad \quad \sigma=t+\int{\frac{dx}{\sqrt{a(x)}}}.
$$
while $\alpha_0$ and $\beta_0$ are arbitrary functions. Therefore, once $h(x,t)$ is given, the linear combination of the solutions of  (\ref{red}) gives the general solution of (\ref{e3}). For instance, if we assume 
\begin{equation}
h=h_0 (x) e^{-k_0 t}  \label{els}
\end{equation}
where $k_0$ is a constant, taking (\ref{albe2}) into account and setting $\alpha_0=\beta_0=0$, by integrating (\ref{red}) we obtain
$$
u_1=a^{-\frac{1}{4}}\left(  h_1 (x) e^{-k_0 \sigma}+u_{1}^{0}(\sigma)\right), \quad \quad  u_2=a^{-\frac{1}{4}}\left(  h_2 (x) e^{-k_0 \xi}+u_{2}^{0}(\xi)\right)  
$$
where $u_{1}^{0}$,  $u_{2}^{0}$ are arbitrary functions, while
$$
h_1 (x)=-\int{\left( \frac{e^{2k_0 \int{\frac{dx}{\sqrt{a}}}}}{\sqrt{a}}\int{h_0 (x) a^{-\frac{1}{4}}e^{-k_0 \int{\frac{dx}{\sqrt{a}}}}dx}\right) dx}, \quad h_2(x)=-\int{\left( \frac{e^{-2k_0 \int{\frac{dx}{\sqrt{a}}}}}{\sqrt{a}} \int{h_0 (x) a^{-\frac{1}{4}}e^{k_0  \int{\frac{dx}{\sqrt{a}}}}dx}\right) dx}
$$
Thus, the general solution of (\ref{e3}), supplemented by (\ref{els}) and  $c(x)$ characterized by (\ref{q2}), is  given by
\begin{equation}
u=\frac{1}{2} (u_1+u_2)=\frac{a^{-\frac{1}{4}}}{2}\left( h_1(x) e^{-k_0 \sigma}+u_{1}^{0}(\sigma)+h_2(x) e^{-k_0 \xi} + u_{2}^{0}(\xi) \right)  \label{s6}
\end{equation}
whatever function $a(x)$ is given.
Furthermore, if $h=0$ and $c=0$, equation (\ref{e3}) assumes the form
\begin{equation}
u_{tt}=\partial_x \left( a(x) u_x \right).   \label{eee}
\end{equation}
In such a case, from (\ref{q2}) we have
\begin{equation}
a(x)=c_0 \, x \sqrt[3]{x}   \label{fi} 
\end{equation}
where $c_0$ is a constant, so that, owing to (\ref{s6}), the general solution of (\ref{eee}) assumes the form
\begin{equation}
u=\frac{1}{\sqrt[3]{x}} \left( u_{1}^{0} (\sigma) + u_{2}^{0} (\xi) \right), \quad \quad \sigma=t+\frac{3}{\sqrt{c_0}} \, \sqrt[3]{x}, \quad \quad \xi=t- \frac{3}{\sqrt{c_0}} \, \sqrt[3]{x}.  \label{sol3}
\end{equation}

\vspace{0,2cm}
\noindent
{\bf Remark 5.} It could be of some interest to notice that in \cite{kap1} a transformation mapping the equation (\ref{eee}) to the Klein Gordon equation
\begin{equation}
v_{tt}=v_{\xi \xi}+ \mu (\xi) v   \label{kg}
\end{equation}
is found. Therefore,  the general solution of (\ref{eee}) characterized by (\ref{sol3}) can be also useful for solving equations like (\ref{kg}).

\vspace{0,2cm}
\noindent
{\ iii)} The equation
\begin{equation}
u_{tt}-u_{xx}=A(x) u_x +B(x,t,u) \label{e4}
\end{equation}
when $A=\frac{\alpha_0}{x}$ and $B=h(x,t)$ was considered in \cite{mohanty}, while when $A=\frac{c_0}{x}$ and $B=-\frac{k_0}{x^2}u$ was studied in \cite{koc}. First, we point out our attention to the case $A=\frac{\alpha_0}{x}$ and $B=h(x,t)$, where $\alpha_0$ denotes  a constant. In the present case we have
\begin{equation}
\gamma=\frac{\alpha_0}{2x}, \quad \quad \eta=-\frac{\alpha_0}{2x} \label{m1}
\end{equation}
while conditions (\ref{ge}) requires
$$
\alpha_0=0 \quad \quad \mbox{or} \quad \quad \alpha_0=2.
$$
Integration of (\ref{al}) and (\ref{be}) gives
\begin{eqnarray}
&&\alpha=(\sigma + \xi)^{-\frac{\alpha_0}{2 }}\left( \int{\frac{h(\xi, \sigma)}{2}(\sigma + \xi)^{\frac{\alpha_0}{2}}d\xi}+\mu (\sigma)\right)  \label{equ1} \\
&&\beta=(\sigma + \xi)^{-\frac{\alpha_0}{2 }}\left( -\int{\frac{h(\xi, \sigma)}{2}(\sigma + \xi)^{\frac{\alpha_0}{2}}d\sigma}+\nu (\xi)\right)  \label{equ2}
\end{eqnarray}
where
$$
\sigma=x-t, \quad \quad \xi=x+t.
$$
while $\mu$ and $\nu$ are arbitrary. Next, by solving (\ref{red}), we find
\begin{eqnarray}
&&u_1=\left( \sigma + \xi \right)^{-\frac{\alpha_0}{2}} \left\{ -\frac{1}{2}\left( \int{\mu(\sigma) d\sigma} +\frac{1}{2} \int{\left( \int{ h(\xi, \sigma) \left(\sigma + \xi \right)^{\frac{\alpha_0}{2}}d\xi}\right) d\sigma} \right) + u_{1}^{0}(\xi) \right\}  \nonumber \\
&&u_2=\left( \sigma + \xi \right)^{-\frac{\alpha_0}{2}} \left\{ \frac{1}{2}\left( \int{\nu (\xi) d\xi} -\frac{1}{2} \int{\left( \int{ h(\xi, \sigma) \left(\sigma + \xi \right)^{\frac{\alpha_0}{2}}d\sigma}\right) d\xi} \right) + u_{2}^{0}(\sigma) \right\} \nonumber
\end{eqnarray}
where the functions $u_{1}^{0}$, $u_{2}^{0}$ are arbitrary. Therefore, the general solution of (\ref{e4}) with  $A=\frac{\alpha_0}{x}$ and $B=h(x,t)$ is given by
$$
u=\frac{1}{2}(u_1+u_2)=\frac{\left( \sigma + \xi \right)^{-\frac{\alpha_0}{2}}}{2} \left( u_{1}^{0}(\xi) + u_{2}^{0} (\sigma) -\frac{1}{2} \int{\left(\int{h(\xi, \sigma) \left( \sigma + \xi \right)^{\frac{\alpha_0}{2}}} d\xi \right)d\sigma}\right)
$$
if $\alpha_0=0$ or $\alpha_0=2$ and where we set, without loss of generality, $\mu=\nu=0$.

\vspace{0.2cm}
\noindent
Now we consider (\ref{e4}) with $A=\frac{c_0}{x}$ and $B=-\frac{k_0}{x^2}u$, where $c_0$ and $k_0$ are constants.  In such a case equation (\ref{e4}) specializes to the Klein-Gordon-Fock (KGF) equation with central symmetry. It results that
\begin{equation}
\gamma=\frac{c_0}{2x}, \quad \quad \eta=-\frac{c_0}{2x} \label{mm1}
\end{equation}
while from (\ref{ge}) we have
\begin{equation}
k_0=\frac{c_0}{2}-\frac{c_0^2}{4}.  \label{v6}
\end{equation}
By solving (\ref{al}) and (\ref{be}) we obtain
$$
\alpha=\mu(\sigma) \left( \sigma + \xi \right)^{-\frac{c_0}{2}}, \quad \quad \beta=\nu(\xi) \left( \sigma + \xi \right)^{-\frac{c_0}{2}}
$$
where
$$
\sigma=x-t, \quad \quad \xi=x+t.
$$
and  $\mu$ and $\nu$ are arbitrary functions. Integration of (\ref{red}) gives
$$
u_1=\left( \sigma +\xi \right)^{-\frac{c_0}{2}} \left( - \int{\frac{\mu(\sigma)}{2} \, d\sigma} +u_{1}^{0}(\xi) \right), \quad \quad u_2=\left( \sigma +\xi \right)^{-\frac{c_0}{2}} \left(  \int{\frac{\nu(\xi)}{2} \, d\xi} +u_{2}^{0}(\sigma) \right)
$$
with $u_{1}^{0}$ and  $u_{2}^{0}$ arbitrary functions. By setting without loss of generality $\mu=\nu=0$, the general solution of (\ref{e4}) when $A=\frac{c_0}{x}$ and $B=-\frac{k_0}{x^2}u$ is given by
$$
u=u_1+u_2=x^{-\frac{c_0}{2}} \left( u_{1}^{0}(\xi)+u_{2}^{0}(\sigma) \right)
$$
provided that condition (\ref{v6}) is satisfied.
 
\vspace{0.2cm}
\noindent
{\it iv)} Finally we consider the equation (\ref{e2}) with $h=-\frac{c_0^2}{4} u+h_0 (x,t)$. It follows that
$$
\gamma=\eta=-\frac{c_0}{2}
$$
while  (\ref{ga}) and (\ref{et}) are identically satisfied. Next, integration of (\ref{al}) and (\ref{be}) gives
\begin{equation}
\alpha=e^{-\frac{c_0}{4}\sigma}\left( \int{\frac{h_0(\xi, \sigma)}{2} \, e^{\frac{c_0}{4}\sigma}d\sigma} + \alpha_0 (\xi) \right), \quad \quad \beta=e^{\frac{c_0}{4}\xi}\left( -\int{\frac{h_0(\xi, \sigma)}{2} \, e^{-\frac{c_0}{4}\xi}d\xi} + \beta_0 (\sigma) \right) \label{bes}
\end{equation}
where
$$
\sigma=x+t, \quad \quad \xi=x-t
$$
while $\alpha_0$ and $\beta_0$ arbitrary. Taking (\ref{bes}) into account, the solution of the equations (\ref{red}) is
$$
u_1=e^{\frac{c_0}{4}\xi} \left( -\int{\frac{\alpha(\xi, \sigma)}{2}\, e^{-\frac{c_0}{4}\xi}d\xi} + u_{1}^{0} (\sigma) \right), \quad \quad u_2=e^{-\frac{c_0}{4}\sigma} \left( \int{\frac{\beta(\xi, \sigma)}{2}\, e^{\frac{c_0}{4}\sigma}d\sigma} + u_{2}^{0} (\xi) \right)
$$
Finally, the general solution of the equation (\ref{e2}) with $h=-\frac{c_0^2}{4} u+h_0(x,t)$ is given by
$$
u=\frac{1}{2}(u_1+u_2)=\frac{1}{2} \left\{ e^{-\frac{c_0}{4}t}\left(e^{\frac{c_0}{4}x} \, u_{1}^{0}(\sigma) +e^{-\frac{c_0}{4}x} \, u_{2}^{0}(\xi) \right)-e^{-\frac{c_0}{2}t} \int{\left(\int{\frac{h_0(\xi, \sigma)}{2}e^{\frac{c_0}{4}\left( \sigma - \xi \right)}d\xi} \right)d\sigma} \right\}
$$
where $u_{1}^{0}(\sigma)$ and $u_{2}^{0}(\xi)$ are arbitrary functions, while, without loss of generality, we set $\alpha_0=\beta_0=0$.

\section{Conclusions}
In this paper we developed a reduction procedure for determining exact solutions of second order hyperbolic equations. The approach considered permits to reduce the integration of a second order equation to that of a first order PDE called intermediate integral. The solutions obtained, apart from their theoretical value, can be also useful for testing numerical integration methods.

We proved that any second order hyperbolic PDE admits two intermediate integrals so that two particular solutions given in terms of one arbitrary function can be calculated. In Theorem $1$ we characterized the compatibility conditions in order that such intermediate integrals exist.

The procedure here developed is particularly useful in the case of linear second order hyperbolic equations. In fact, in such a case, the linear combination of the solutions of the two intermediate integrals gives the solution of the second order governing equation in terms of two arbitrary functions. Therefore any initial value problems can be solved.  In Theorem $2$ we characterized the class of the linear second order hyperbolic PDEs for which it is possible to obtain the general solution by means of the procedure here considered.

The reduction method here developed should be applied, in principle, to any second order or higher order PDE but, as far as we know, such a procedure has been applied only for hyperbolic equations.  Therefore it could be of some interest to look for intemediate integrals, for instance, for parabolic reaction-diffusion equations. This is our aim for future research.

\vspace{0,3cm}

\subsection*{Acknowledgements} The authors thank the financial support of GNFM of the Istituto Nazionale di Alta Matematica and of PRIN\_202248TY47\_003,\textquotedblleft Modelling complex biOlogical systeMs for biofuEl productioN and sTorAge: 
mathematics meets green industry (MOMENTA)  \textquotedblright .

\end{document}